# Charge-stripe order in the electronic ferroelectric $LuFe_2O_4$

Y. Zhang, H. X. Yang, C. Ma, H. F. Tian & J. Q. Li *

Beijing National Laboratory for Condensed Matter Physics, Institute of Physics, Chinese Academy of Sciences, Beijing 100080, P. R. China.

The structural features of the charge ordering states in $LuFe_2O_4$ are characterized by in-situ cooling TEM observations from 300K down to 20K. Two distinctive structural modulations, a major $\mathbf{q_1}= (1/3, 1/3, 2)$ and a weak $\mathbf{q_2}=\mathbf{q_1}/10 + (0, 0, 3/2)$, have been well determined at the temperature of ~20K. Systematic analysis demonstrates that the charges at low temperatures are well crystallized in a charge stripe phase, in which the charge density wave behaviors in a non-sinusoidal fashion resulting in elemental electric dipoles for ferroelectricity. It is also noted that the charge ordering and ferroelectric domains often change markedly with lowering temperatures and yields a rich variety of structural phenomena.

Understanding charge ordering (CO) and the remarkable electronic ferroelectricity in the charge frustrated $RFe_2O_4$ (R = rare-earth elements) system is regarded as an important issue from both academic and technological points of view[1,2]. The origin of ferroelectricity in this kind of material is fundamentally correlated with the CO transition arising from strong electron correlation[3-5]; the giant room-temperature magnetodielecric response[6] as demonstrated in $LuFe_2O_4$ could give rise to a new generation of multifunctional devices for microelectronics[7, 8]. $RFe_2O_4$ consists of two layers, a hexagonal double layer of Fe ions, with an average valence of $Fe^{2.5+}$, which are sandwiched by a thick $R_2O_3$ layer[9]. The spin/charge behavior in the double Fe-layers directly affect the magnetic features and, in particular, the ferroelectricity[10,11]. In previous literature, though certain attempts have been made to clarify the correlation between the $Fe^{2+}/Fe^{3+}$ order and the ferroelectricity in $RFe_2O_4$, the details of the CO states are not known at this time[6,12-15]. In this letter, we will report on the clear CO modulations as revealed by in-situ cooling TEM observations in $LuFe_2O_4$. It is demonstrated for the first time that the charges in the ground state are well crystallized in a charge stripe phase which can properly explain the remarkable electronic ferroelectricity in $LuFe_2O_4$.

The crystal structure, ferroelectricity and magnetic properties of all $LuFe_2O_4$ samples used in present study have been well characterized by a variety of experimental measurements. Similar results have been obtained to those described in previous literature[13,16,17]. The specimens for TEM observations were prepared by polishing, dimpling and subsequently ion milling. Microstructure analysis was performed on Tecnai F20 and H-9000NA transmission electron microscopes both equipped with low temperature holders.

In order to clearly view the structural features in correlation with the CO transitions in $LuFe_2O_4$ materials, we have carried out a number of in-situ TEM observations along several relevant crystallographic directions. Our experimental results suggest that the fundamental properties of the superstructures in this charge-frustrated system depend considerably on temperature, and, in general, the sharp superstructure spots become visible below 50 K. Fig. 1a and b show respectively the [1-10] and [001] zone-axis electron diffraction patterns taken at around 20 K, demonstrating the presence of clear superstructure spots following the main diffraction spots. The main spots with relatively strong intensity in both diffraction patterns can be well indexed on the expected hexagonal unit cell with lattice parameters of a=3.444 Å, and c=25.259 Å (space group of R-3m). On the other hand, the superstructure spots can be assigned respectively to two structural modulations ($q_1$ and $q_2$) as clearly illustrated in the schematic pattern of Fig. 1c. Fig. 1b represents a typical electron diffraction pattern taken along the direction of slightly off the c*-direction, demonstrating that the weak spots of $q_1$ modulation are visible in three symmetric <110>-directions. Our careful analysis suggests that the $q_1$-modulation, with a wave vector of $q_1$= (1/3, 1/3, 2), can be directly interpreted by a $Fe^{2+}/Fe^{3+}$ order as discussed in the following context. While the $q_2$ modulation in general is much weaker than the $q_1$ modulation, its average wave vectors at 20 K can be roughly written as $q_2$=(0 0 3/2)+ $q_1$/10. Further experimental measurements indicate that the $q_2$ modulation also depends on the local defective structures. Hence, in the following context, we will devote our main attention on investigations of the $q_1$ modulation.

Fig. 1d shows a structural model illustrating ionic ordering of $Fe^{2+}$, $Fe^{2.5+}$, and $Fe^{3+}$ for the $q_1$ modulation. This structural model can be used to explain the main experimental results obtained at low temperatures. In addition to the ionic ordering within the Fe double-layers, a relative shift between neighboring Fe layers in the CO patterns is clearly visible in the ground state. These facts suggest that the $q_1$-modulation in $LuFe_2O_4$ corresponds with a well-defined three-dimensional ionic ordering. Fig. 1e displays a theoretically simulated pattern for $LuFe_2O_4$ considering qualitatively local structural distortion as suggested in ref.1 and the difference of scattering factors between $Fe^{2+}$ and $Fe^{3+}$ in the CO state[18], it is clear that the systematic positions of the satellite spots in this pattern are in good agreement with those shown in Fig. 1a.

The origin of ferroelectricity in $LuFe_2O_4$ materials is considered to be essentially in correlation with the low-temperature CO. We firstly make a systematical analysis on the possible electric polarization corresponding to the $q_1$-modulation. Fig. 2a and b shows the schematic charge patterns in accordance with our experimental observations. It is remarkable that positive charges ($Fe^{3+}$ sites) and negative charges ($Fe^{2+}$ sites) are actually crystallized in parallel charge stripes along the view direction. This charge stripe phase shows up a clear monoclinic feature with an evident electric polarity. In order to facilitate the analysis on the electric polarization, we can also characterize this charge concentration in this stripe phase as a charge-density wave (CDW) in association with the $q_1$-modulation. This CDW with the periodicity of about 5Å along the <116>-direction does not behave in a simple sinusoidal fashion, but is strongly affected by charge frustration as illustrated in Fig.

2c. It is easily visible that the average centers of $Fe^{2+}$ (negative) and $Fe^{3+}$ (positive) planes have a clear relative shift and this directly results in a local electric polarization along the $q_1$ direction. From the point of view of ferroelectricity, this kind of charge configuration has notable similarity with the coherent arrangement of electric dipoles discussed commonly in conventional ferroelectric materials[19,20]. We therefore conclude that the $LuFe_2O_4$ crystal has a ferroelectric CO ground state in which the coherent arrangements of electric dipoles are realized by charge stripe order. It is noted that the parent phase of $LuFe_2O_4$ has the rhombohedral symmetry with the space group of R-3m, hence, the CO modulation, together with the local electric polarization ($P_1$), could evenly appear in three crystallographically equivalent <116>-directions around the rhombohedra axis, as clearly illustrated in Fig. 2d. Hence, the resultant spontaneous electric polarization (P) from combination of P(1), P(2) and P(3) occurs along the c-axis direction in $LuFe_2O_4$. This conclusion is in good agreement with the major experimental data reported in previous literature[10,18].

In addition to the low-temperature CO ground state, the temperature dependence of the CO modulations is also an important issue for the understanding of the structural and physical properties in this charge frustrated system. Our TEM observations reveal a variety of interesting microstructure features from room temperature down to 20 K. Fig. 3a shows a series of electron diffraction patterns illustrating superstructure features at different temperatures. In order to facilitate comparison, all the diffraction patterns, especially the superstructure features, are schematically illustrated in the right column of Fig. 3a. The most striking features demonstrated in these diffraction patterns are the clear twinning structure for the charge stripe order and alternation of the CO coherent length along the c-axis direction. Fig. 3b shows an electron diffraction pattern obtained at ~100K, demonstrating the recognizable twinning feature for both $q_1$ and $q_2$ modulations as schematically illustrated in Fig. 3c. The twinning plane in general is parallel with the basal a-b plane. Careful measurements also suggest that the thickness of the twinned lamellae decreases progressively with the increase of temperature from 20 K to 300 K, as a result, all superstructure reflections change from clear spots at 20 K to diffuse streaks at room temperature. In-situ TEM observations on the well-defined CO state show that the average thickness of the CO lamellae is as large as about 50-500 nm at around 20 K so that clear superstructure spots can be easily obtained in selected area diffraction patterns as demonstrated in Fig. 1a. Fig. 4a shows a schematic structural model illustrating the CO patterns corresponding with the (001) twinning structure in the $LuFe_2O_4$ materials. It is important to point out that this kind of CO twinning corresponds directly with the 180° ferroelectric domain structure. Fig. 4b shows a schematic model for the typical 180° ferroelectric domain in $LuFe_2O_4$. In fact, a clear view of this kind of ferroelectric domains can be achieved along [1$\bar{1}$0] zone axis direction and is recognizable as the twinning relationship between adjacent CO lamellae.

Fig. 4c shows a dark-field TEM image obtained at about 100K using the superstructure reflection of $q_1$ modulation, revealing the notable contrast alternations from the micro-domain structure. The average thickness of the twinning lamellae (ferroelectric domains) in this area is estimated in the range of from 10 nm to 30 nm. Diffraction observations between 200 K and 300 K always reveal the superstructure reflections as diffuse streaks on (h/3, h/3, l) lines along c*-direction. Moreover, the superstructure reflections always contain notable contributions from the CO twinning, which often results in a zig-zag type of superstructure streaks as noted in previous publications[1,15]. The average thickness of the CO lamellae at room temperature is estimated to be smaller than 5 nm. Fig. 4d shows a high resolution TEM image illustrating the CO ordered state in $LuFe_2O_4$ at room temperature. The short range ordered states can be clearly seen in the marked typical areas. It should be mentioned that the coherence length of the CO states, as well as the ferroelectric structure, has notable anisotropic features in this layered system, e.g. the coherence length at room temperature as estimated from TEM images is about 20-30 nm within the a-b plane and about 1-3 nm along the c-axis direction, which is in good agreement with the data obtained from diffraction.

In conclusion, the measurements reported here show for the first time that the charges in $LuFe_2O_4$ are crystallized in ordered stripes at low temperatures, and that the charge concentration in this charge stripe phase can be characterized by a non-sinusoidal CDW which gives rise to a clear electric polarization. The remarkable temperature dependence of the CO modulation as revealed by in-situ TEM observations suggests the presence of substantial dynamic features in this frustrated system. These facts provide the essential elements needed for understanding the mechanism of electric ferroelectricity in these layered materials.

**Acknowledgements** We thank Prof. C.J. Lu and R.I. Walton for their helpful discussions. This work was supported by the 'Outstanding Youth Fund' which is granted by the National Natural Science Foundation of China and by the 973 project (No: 2006CB601001) which is granted by the Ministry of Science and Technology of China.

Figure captions

**Figure 1 Superstructure modulations in LuFe$_2$O$_4$ observed at about 20K.** Electron diffraction patterns taken (a) along the [1-10] zone-axis direction and (b) slightly off the **c** axis direction, showing clear superstructure spots. (c) Schematic representation of Fig. 1a illustrating the **q**$_1$ and **q**$_2$ modulations. (d) A structural model for the ionic order of Fe$^{2+}$ and Fe$^{3+}$. (e) A theoretical diffraction pattern showing the satellite spots corresponding with the **q**$_1$ modulation.

**Figure 2 Charge ordering and spontaneous polarization in LuFe$_2$O$_4$.** (a) and (b) Structural models schematically illustrating the ordered charge stripes in the ground state. (c) The CDW in non-sinusoidal fashion along <116>-direction, corresponding with **q**$_1$-modulation. (d) A schematic illustration for electric polarizations in LuFe$_2$O$_4$ crystal, the spontaneous polarization is along the c –direction.

**Figure 3 Evolution of charge ordering with lowering temperature.** (a) A series of electron-diffraction patterns and (b) schematic illustrations directly exhibiting the superstructure features at different temperatures. (c) Diffraction patterns and (d) schematic illustration demonstrating the twinning structure of the charge stripes.

**Figure 4 Relationship between ferroelectric domains and CO twinning.** (a) A structural model for twins in LuFe$_2$O$_4$. (b) A schematic illustration for the 180$^\text{o}$ ferroelectric domains corresponding with the CO twinning. (c) Dark-field TEM images at the temperature of 100K, the stripe-like twinning lamellae (180$^\text{o}$ ferroelectric domains) are identified. (d) HREM image using the spots from **q**$_1$-modulation clearly exhibiting the short-rang ordered state for the CO state at 300 K.

Figure 1

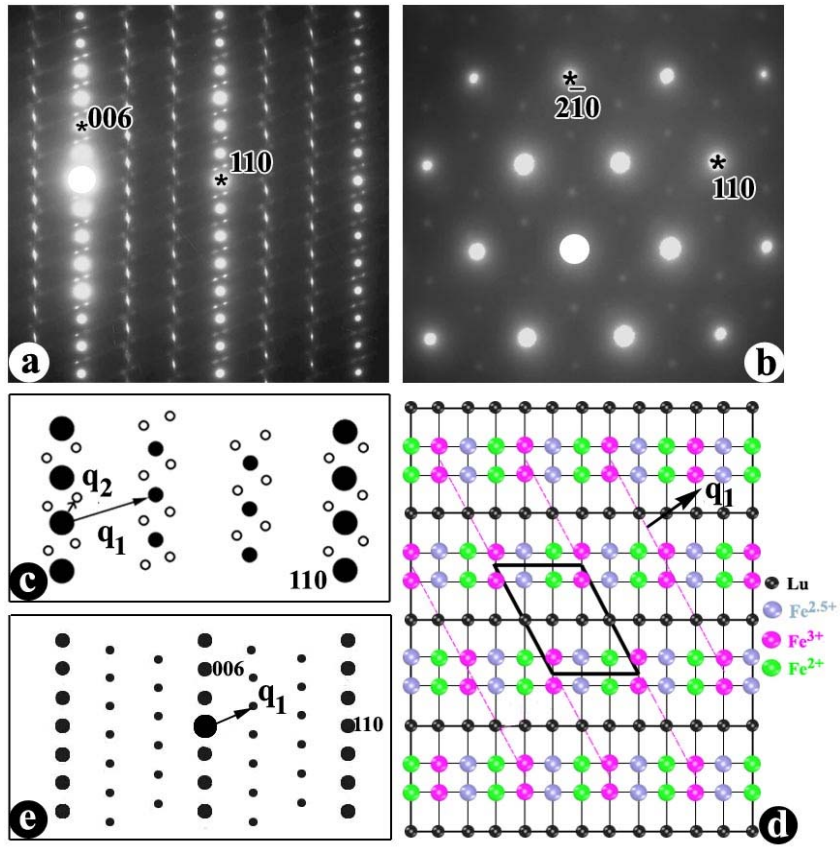

Figure 2

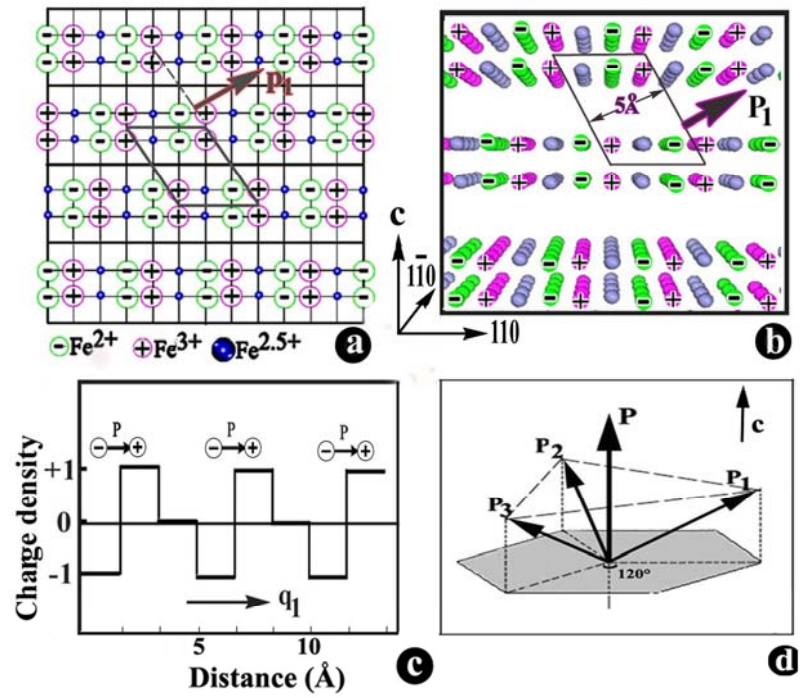

Figure 3

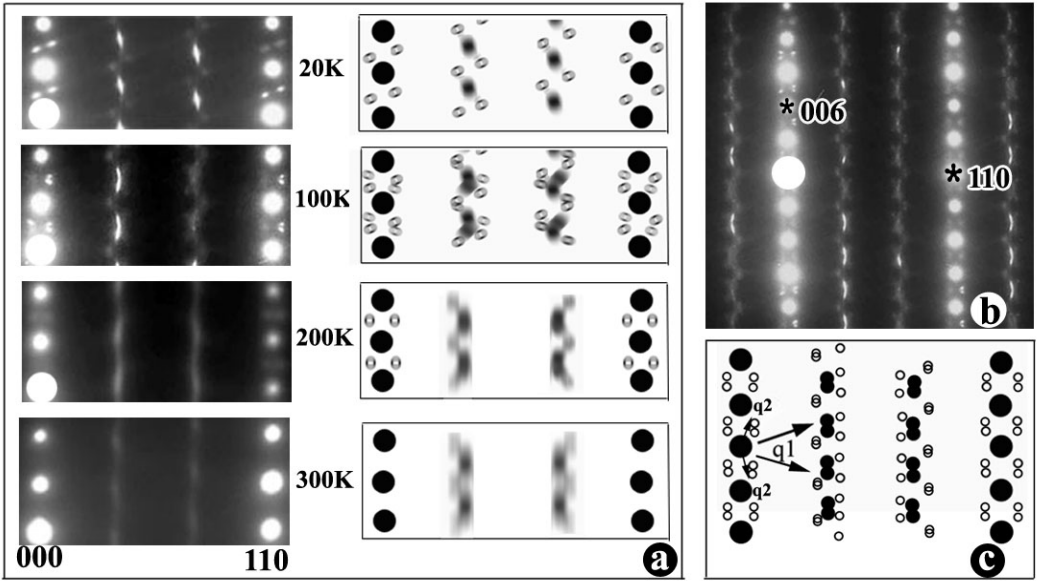

Figure 4

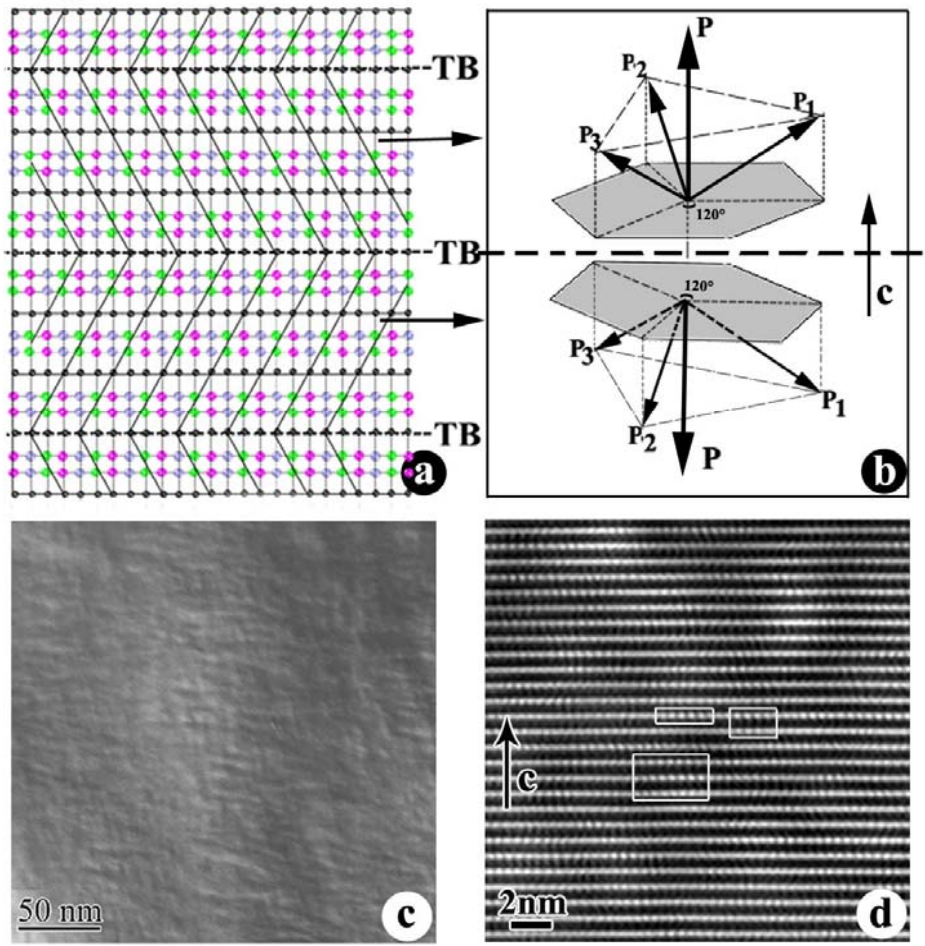